# Catalytic mechanism of LENR in quasicrystals based on localized anharmonic vibrations and phasons

[#] Volodymyr Dubinko [1], Denis Laptev [2], Klee Irwin [3],
[1] NSC "Kharkov Institute of Physics and Technology", Ukraine
[2] B. Verkin Institute for Low Temperature Physics and Engineering, Ukraine
[3] Quantum Gravity Research, Los Angeles, USA
E-mail: vdubinko@hotmail.com

**Abstract**

Quasicrystals (QCs) are a novel form of matter, which are neither crystalline nor amorphous. Among many surprising properties of QCs is their high catalytic activity. We propose a mechanism explaining this peculiarity based on unusual dynamics of atoms at special sites in QCs, namely, *localized anharmonic vibrations* (LAVs) and *phasons*. In the former case, one deals with a large amplitude (~ fractions of an angstrom) time-periodic oscillations of a small group of atoms around their stable positions in the lattice, known also as *discrete breathers*, which can be excited in regular crystals as well as in QCs. On the other hand, phasons are a specific property of QCs, which are represented by *very* large amplitude (~angstrom) oscillations of atoms between two quasi-stable positions determined by the geometry of a QC. Large amplitude atomic motion in LAVs and phasons results in *time-periodic driving* of adjacent potential wells occupied by hydrogen ions (protons or deuterons) in case of hydrogenated QCs. This driving may result in the increase of amplitude and energy of *zero-point vibrations* (ZPV). Based on that, we demonstrate a drastic increase of the D-D or D-H fusion rate with increasing number of modulation periods evaluated in the framework of Schwinger model, which takes into account suppression of the Coulomb barrier due to lattice vibrations.

In this context, we present numerical solution of Schrodinger equation for a particle in a *non-stationary* double well potential, which is driven time-periodically imitating the action of a LAV or phason. We show that the rate of tunneling of the particle through the potential barrier separating the wells is enhanced drastically by the driving, and it increases strongly with increasing amplitude of the driving. These results support the concept *of nuclear catalysis* in QCs that can take place at special sites provided by their inherent topology. Experimental verification of this hypothesis can open the new ways towards *engineering of nuclear active environment* based on the QC catalytic properties.

___________________________________________________________________________________

# Content

## 1. Introduction

The tunneling through the Coulomb potential barrier during the interaction of charged particles presents a major problem for the explanation of low energy nuclear reactions (LENR) observed in solids [1-3]. Corrections to the cross section of the fusion due to the screening effect of atomic electrons result in the so-called "screening potential", which is far too weak to explain LENR observed at temperatures below the melting point of solids. Nobel laureate Julian Schwinger proposed that a substantial suppression of the Coulomb barrier may be possible at the expense of *lattice vibrations* [4, 5]. The fusion rate of deuteron-deuteron or proton-deuteron oscillating in adjacent lattice sites of a metal hydride, according to the Schwinger model, is about $10^{-30}$ s$^{-1}$ [6], which is huge as compared to the conventional evaluation by the Gamov tunnel factor (~ $10^{-2760}$). However, even this is too low to explain the observed excess heat generated e.g. in Pd cathode under D$_2$O electrolysis. The fusion rate by Schwinger is extremely sensitive to the amplitude of *zero-point vibrations* (ZPV) of the interacting ions, which has been shown to increase under the action of time-periodic driving of the harmonic potential well width [6]. Such a driving can be realized in the vicinity of *localized anharmonic vibrations* (LAVs) defined as large amplitude (~ fractions of an angstrom) time-periodic vibrations of a small group of atoms around their stable positions in the lattice. A sub-class of LAV, known as *discrete breathers*, can be excited in regular crystals by heating [1-3, 7] or irradiation by fast particles [8]. Based on that, a drastic increase of the D-D or D-H fusion rate with increasing number of driving periods has been demonstrated in the framework of the modified Schwinger model [6, 8].

One of the most important practical recommendations of the new LENR concept is to look for the *nuclear active environment* (NAE), which is enriched with nuclear active sites, such as the LAV sites. In this context, a striking *site selectiveness* of LAV formation in disordered structures [9] allows one to suggest that their concentration in quasicrystals (QCs) may be very high as compared to regular crystals where discrete breathers arise homogeneously, and their activation energy is relatively high. Direct experimental observations [10] have shown that in the decagonal quasicrystal Al$_{72}$Ni$_{20}$Co$_8$, *mean-square thermal vibration amplitude* of the atoms at special sites substantially exceeds the mean value, and the difference increases with temperature. This might be the first experimental observation of LAV, which has shown that they are arranged in just a few nm from each other, so that their average concentration was about $10^{20}$ per cubic cm that is many orders of magnitude higher than one could expect to find in periodic crystals [1-3, 7]. Therefore, in this case, one deals with a kind of *'organized disorder'* that stimulates formation of LAV, which may explain a strong catalytic activity of quasicrystals [11].

In addition to the enhanced susceptibility to the LAV generation, QCs exhibit unique dynamic patterns called *phasons*, which are represented by *very* large amplitude (~angstrom) quasi time-periodic oscillations of atoms between two quasi-stable positions determined by the geometry of a QC. It is natural to expect that the driving effect of phasons can exceed that of LAVs due to the larger oscillation amplitude in phasons. The main goal of the present paper is to develop this concept to the level of a quantitative comparison between the driving/catalytic action of LAVs and phasons, which could be used to suggest some practical ways of catalyzing LENR.

The paper is organized as follows. In the next section, the Schwinger model [4, 5] and its extension [6] are shortly reviewed to demonstrate an importance of time-periodic driving of potential wells in the LENR triggering.

In section 3, we extend our analysis beyond the model case of infinite harmonic potential (the tunneling from which is impossible) and obtain numerical solution of Schrodinger equation for a particle in a *non-stationary* double well potential, which is driven time-periodically imitating the action of a LAV or phason. We show that the rate of tunneling of the particle through the potential barrier separating the wells is enhanced drastically by the driving, and it increases strongly with increasing amplitude of the driving.

In section 4, we present some examples of dynamical patterns in QCs and their clusters and discuss the ways of experimental verification of the proposed concept. The summary and outlook is given in section 5.

**2. Schwinger model of LENR in an atomic lattice modified with account of time-periodic driving**

According to Schwinger [4], the effective potential of the deuteron-deuteron (D-D) or proton-deuteron (P-D) interactions is modified due to averaging $_0\langle\ \rangle_0$ related to their *zero-point vibrations* (ZPV) in adjacent harmonic potential wells, where $_0\langle\ \rangle_0$ symbolizes the phonon vacuum state. It means that nuclei in the lattice act not like point-like charges, but rather (similar to electrons) they are "smeared out" due to quantum oscillations in the harmonic potential wells near the equilibrium positions. The resulting effective Coulomb interaction potential $_0\langle V_c(r) \rangle_0$ between a proton and a neighboring ion at a distance *r* can be written, according to [4] as

$$_0\langle V_c(r) \rangle_0 = \frac{Ze^2}{r}\sqrt{\frac{2}{\pi}} \int_0^{r/\Lambda_0} dx \exp\left(-\tfrac{1}{2}x^2\right) \approx \begin{cases} r \gg \Lambda_0 : \dfrac{e^2}{r} \\ r \ll \Lambda_0 : \left(\dfrac{2}{\pi}\right)^{1/2} \dfrac{e^2}{\Lambda_0} \end{cases}, \qquad (1)$$

where *Z* is the atomic number of the ion, *e* is the electron charge, $\Lambda_0 = (\hbar/2m\omega_0)^{1/2}$ is the ZPV amplitude, $\hbar$ is the Plank constant, *m* is the proton mass, and $\omega_0$ is the angular frequency of the harmonic potential. A typical value of $\Lambda_0 \sim 0.1$ Å, which means that the effective repulsion potential is saturated at ~ *several hundred eV* as compared to *several hundred keV* for the unscreened Coulomb interaction. Schwinger estimated the rate of fusion as the rate of transition out of the phonon vacuum state, which is reciprocal of the mean lifetime T$_0$ of the vacuum state, which can be expressed via the main nuclear and atomic parameters of the system [5, 6]:

$$\frac{1}{T_0} \sim 2\pi\omega_0 \left(\frac{2\pi\hbar\omega_0}{E_{nucl}}\right)^{\frac{1}{2}} \left(\frac{r_{nucl}}{\Lambda_0}\right)^3 \exp\left[-\frac{1}{2}\left(\frac{R_0}{\Lambda_0}\right)^2\right] \qquad (2)$$

where $E_{nucl}$ is the nuclear energy released in the fusion, which is transferred to the lattice producing phonons (*that explains the absence of harmful radiation in LENR*), $r_{nucl}$ is the nuclear radius, $R_0$ is the equilibrium distance between the nuclei in the lattice.

For D-D => He$^4$ fusion in PdD lattice, the mass difference $E_{nucl}$ = 23.8 MeV. Assuming $r_{nucl} = 3\times 10^{-5}$ Å, $\Lambda_0$ = 0.1 Å (corresponding to $\omega_0$ = 320 THz) and $R_0$ =0.94 Å as the equilibrium spacing of two deuterons placed in *one site* in a hypothetical PdD$_2$ lattice, Schwinger estimated the fusion rate to be ~ 10$^{-19}$ s$^{-1}$ [5]. For a more realistic situation, with two deuterons in *two adjacent sites* of the PdD lattice, one has $R_0$ =2.9 Å. Even assuming a lower value of $\omega_0$ = 50 THz corresponding to larger $\Lambda_0$ = 0.25 Å [6], eq. (2) will results in the fusion rate of ~ 10$^{-30}$ s$^{-1}$, which is too low to explain the observed excess heat generated in Pd cathode under D$_2$O electrolysis.

The above estimate is valid for the fusion rate between D-D or D-H ions in regular lattice sites. However, the ZPV amplitude can be increased locally under time-periodic modulation of the potential well width (that determines its eigenfrequency) at a frequency that exceeds the eigenfrequency by a factor of ~2 (*the parametric regime*). Such regime can be realized for a hydrogen or deuterium atom in metal hydrides/deuterides, such as NiH or PdD, in the vicinity of LAV [2, 3]. Under parametric modulation, ZPV amplitude increases exponentially fast (Fig. 1a) with increasing number of oscillation periods $N = \omega_0 t/2\pi$ [6]:

$$\Lambda_N = \Lambda_0 \sqrt{\cosh(g_\omega \pi N)}, \quad \Lambda_0 = \sqrt{\frac{\hbar}{2m\omega_0}}, \qquad (3)$$

where $g_\omega \ll 1$ is the amplitude of parametric modulation, which is determined by the amplitude of LAV. For example, $g_\omega$ = 0.1 corresponds to the LAV amplitude of ~ 0.3 Å in the PdD lattice with $R_0$ =2.9 Å, which is confirmed by molecular dynamic simulations of gap discrete breathers in NaCl type crystals [7]. Substituting eq. (3) into the Schwinger eq. (2) one obtains a drastic enhancement of the fusion rate with increasing number of oscillation periods N (Fig. 1b):

$$\frac{1}{T_N} \sim 2\pi\omega_0 \left(\frac{2\pi\hbar\omega_0}{E_{nucl}}\right)^{\frac{1}{2}} \left(\frac{r_{nucl}}{\Lambda_N}\right)^3 \exp\left[-\frac{1}{2}\left(\frac{R_0}{\Lambda_N}\right)^2\right], \tag{4}$$

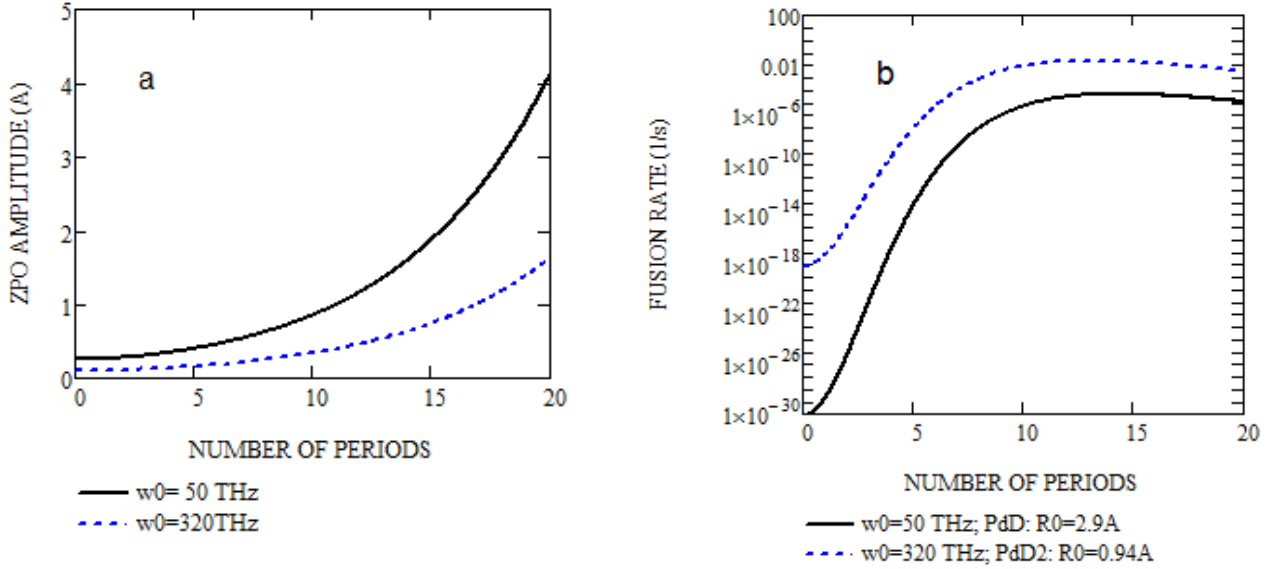

Figure 1. (**a**) Zero-point vibration amplitude of deuterium ions vs. *N* in the *parametric regime* [6] for different $\omega_0$ according to eq. (3) at $g = 0.1$. (**b**) D-D fusion rate D-D => He$^4$ + 23.8 MeV in PdD lattice according to eq. (4) for deuterium ions in PdD lattice oscillating near equilibrium positions in one site ($R_0$ =0.94 Å) or in two neighboring lattice sites ($R_0$ =2.9 Å).

The parametric driving considered above requires rather special conditions similar to those in gap breathers in diatomic crystals [7], while in many other systems, e.g. in metals [12], oscillations of atoms in a discrete breather have different amplitudes but the same frequency. This case is more close to the driving of the potential well *positions* with the frequency equal to the potential eigenfrequency. Such driving does not increase the ZPV amplitude since the wave packet dispersion remains constant, however, the mean oscillation energy grows with time as [13]:

$$\langle E \rangle = \frac{\hbar\omega_0}{2} + \frac{g_x^2 \hbar\omega_0}{16}\left[\omega_0^2 t^2 + \omega_0 t \sin 2\omega_0 t + \sin^2 \omega_0 t\right], \tag{5}$$

where $g_x$ is the relative amplitude of the position driving. Accordingly, one could expect an acceleration of the escape from a potential well of a finite depth similar to the parametric driving.

In reality, one is interested in the effect of potential well driving on the *tunneling* through the barrier of finite height between the wells as a function of the driving frequency and strength (amplitude). Analytical solution of the *non-stationary* Schrödinger equation even for the simplest case of a double well potential cannot be obtained. In the following section, we will analyze a *numerical solution* of Schrödinger equation for a particle in a double well potential, which is driven time-periodically imitating the action of a LAV or phason.

### 3. Tunneling in a periodically-driven double well potential

Consider Schrödinger equation for a wave function $\psi(x,t)$ of a particle with a mass *m* in the non-stationary double-well potential $V(x,t)$:

$$i\hbar \frac{\partial}{\partial t}\psi(x,t) = -\frac{\hbar^2}{2m}\frac{\partial^2}{\partial x^2}\psi(x,t) + V(x,t)\psi(x,t), \qquad (6)$$

$$V(x,t) = \frac{m\omega_0^2}{2}\left[\frac{a(t)}{x_0^2}x^4 - b(t)x^2\right], \qquad x_0 = \sqrt{\frac{\hbar}{m\omega_0}} \qquad (7)$$

where $a(t)$ and $b(t)$ are the dimensionless parameters that determine the form and the driving mode of the potential shown in Fig.2.

$$a(t) = \frac{1}{2\sqrt{\alpha}}\left[\alpha - \beta\cos(\Omega t)\right], \; b(t) = \frac{1}{2\sqrt{\alpha}}\sqrt{\alpha - \beta\cos(\Omega t)}, \qquad (8)$$

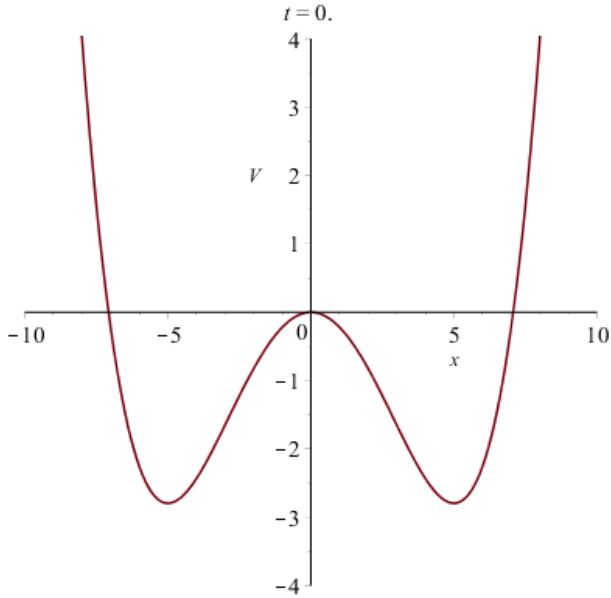

Figure 2. Double-well potential given by eq. (7) at $\alpha = 0.0005$, $\beta = 0.0001$, which correspond to the ratio of the potential depth to ZPV energy given by $\frac{1}{8\sqrt{\alpha}} \approx 5.6$

$\Omega$ is the driving frequency of the eigenfrequencies $\omega_{eigen}$ and positions $x_{min}$ of the potential wells in the vicinity of the minima given by

$$\frac{\omega_{eigen}}{\omega_0} = \sqrt{2b} = \sqrt[4]{1 - \frac{\beta}{\alpha}\cos(2\omega_0 t)} \approx \left[1 - \frac{\beta}{4\alpha}\cos(2\omega_0 t)\right], \qquad g_\omega = \frac{\beta}{2\alpha} \ll 1, \; (9)$$

$$\frac{x_{min}}{x_0} = \pm\sqrt{\frac{b}{2a}} = \pm\frac{1}{\sqrt{2}}\frac{1}{\sqrt[4]{\alpha - \beta\cos(2\omega_0 t)}} = \frac{1}{\sqrt{2}}\frac{1}{\alpha^{1/4}\sqrt[4]{1 - \frac{\beta}{\alpha}\cos(2\omega_0 t)}},$$

$$\approx \frac{x_{min}(0)}{x_0}\left(1 + \frac{\beta}{4\alpha}\cos(2\omega_0 t)\right), \; g_x \equiv \frac{\beta}{4\alpha} \ll 1 \qquad (10)$$

From eqs (9), (10) it follows that the driving under consideration [https://www.dropbox.com/s/eczwm6ny8f0939t/Potential%20driving.gif?dl=0] results in a *simultaneous* time-periodic modulation of the potential well *positions* and *eigenfrequencies* with amplitudes $g_x$ and $g_\omega$, respectively. Therefore, we are dealing here with a synergetic effect of the two mechanisms considered separately for a harmonic oscillator in the previous section and in ref. [13].

Initial state of the system is described by a wave function of the Gaussian form placed near the first energy minimum (Fig. 2a):

$$\psi(x, t_0 = 0) = \frac{1}{\sqrt[4]{\pi x_0^2}} \exp\left(-\frac{(x - x_{\min})_0^2}{2x_0^2}\right), \qquad (11)$$

The probability distribution of finding the particle at the point x is given by $\rho(x, t_0 = 0) = |\psi(x, t_0 = 0)|^2$, which is shown in Fig. 3b. It can be seen that the probability density is concentrated at $x_{\min} \approx 4.73$, which means the particle spends most of its time at the bottom of the potential well.

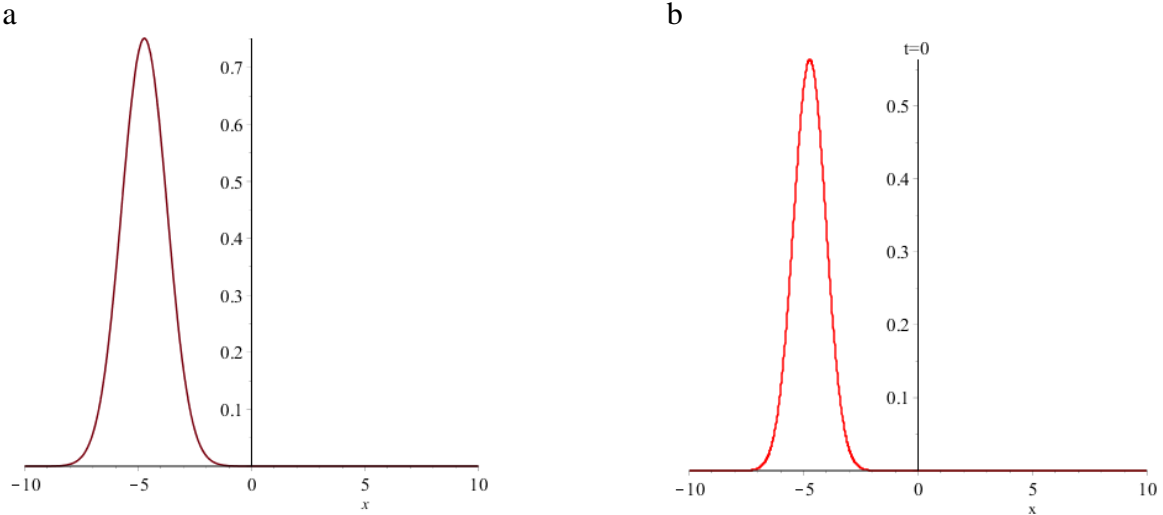

Figure 3. (a) Initial wave function $\psi(x, t_0 = 0)$ and (b) the probability distribution to find the particle at the point x: $\rho(x, t_0 = 0) = |\psi(x, t_0 = 0)|^2$ in the left potential well shown in Fig. 1.

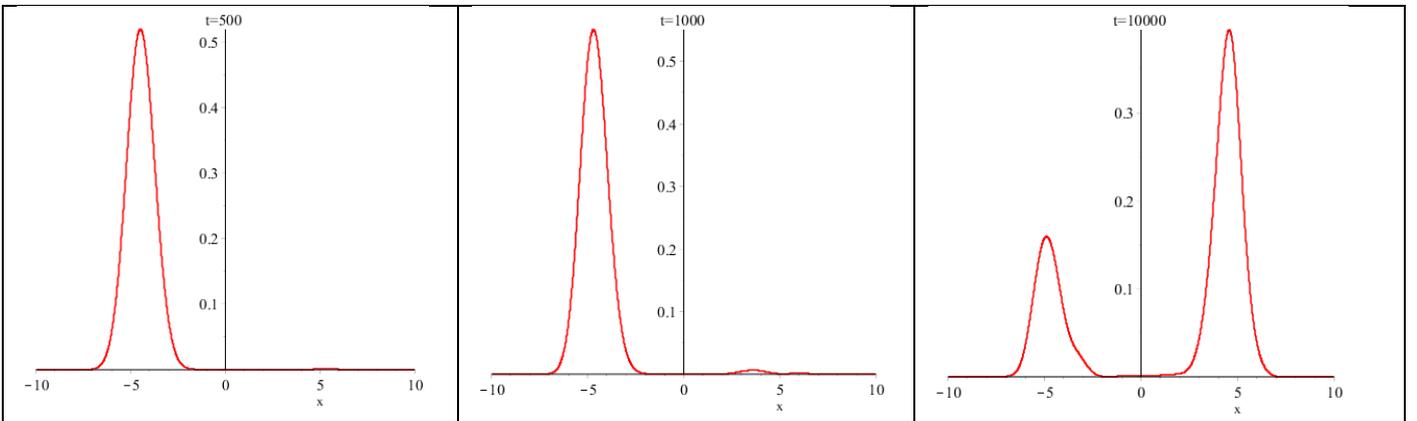

Figure 4. The probability distribution of the particle at different moments of time $t = 2\pi/\omega_{eigen}$ in *stationary potential wells*: $\alpha = 0.0005$; $\beta = 0$.

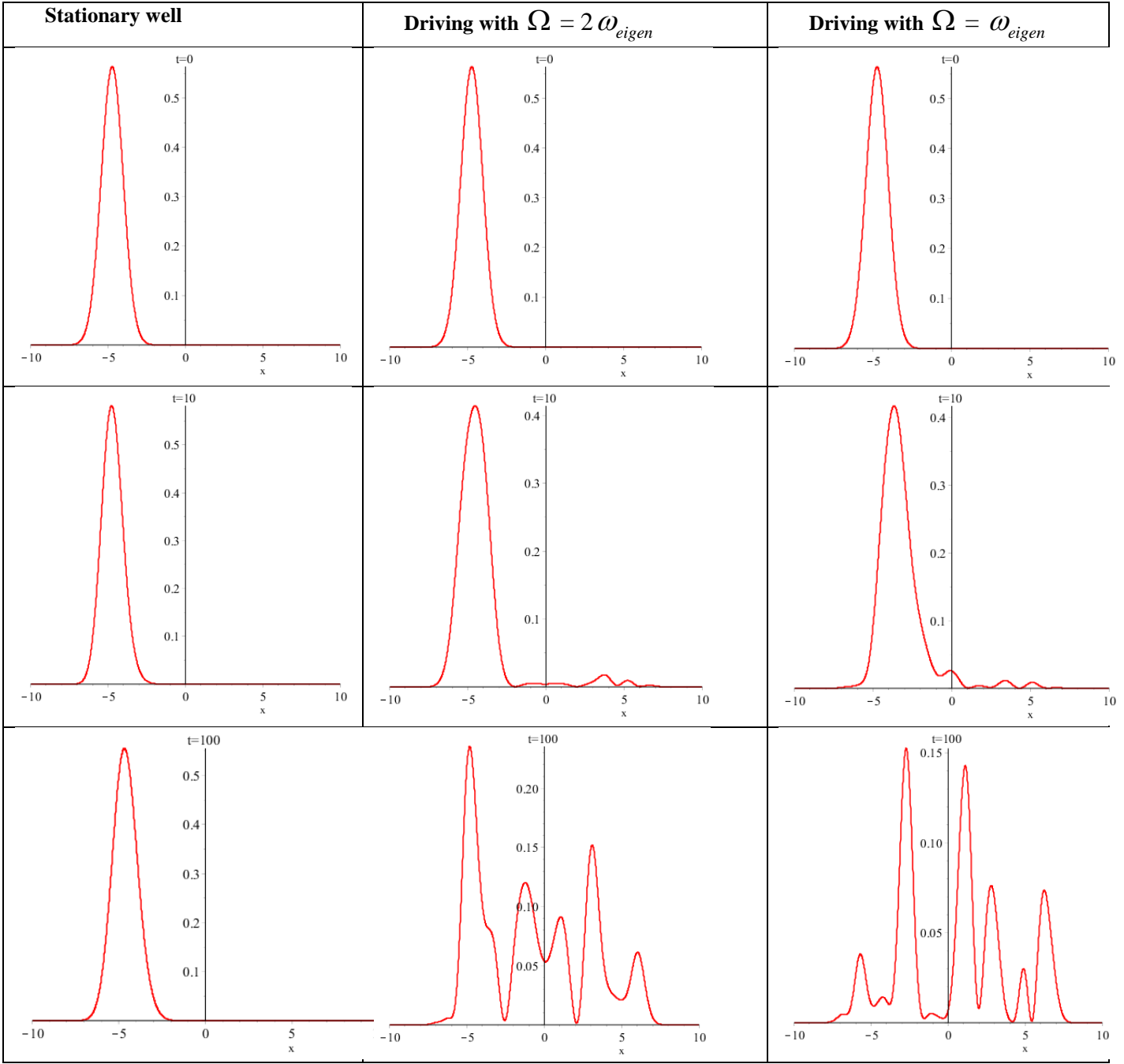

Figure 5. The probability distribution of the particle at different moments of time $t = 2\pi/\omega_{eigen}$ in stationary potential wells ($\alpha = 0.0005$; $\beta = 0$) and under the potential driving ($\alpha = 0.0005$; $\beta = 0.0001$) corresponding to $g_\omega = \beta/2\alpha = 0.1$; $g_x = \beta/4\alpha = 0.05$. The driving frequency $\Omega$ is indicated in the figure.

At the selected parameters, the potential depth to ZPV energy ratio is given by $1/8\sqrt{\alpha} \approx 5.6$, which is a typical ratio for solid state chemical reactions. It means that the particle energy is 5.6 times lower than the energy required to 'jump' over the barrier into another well. The mean time of tunneling through the barrier from a *stationary* potential well is very large as can be seen from Fig. 4 showing the probability distribution of the particle at different moments of time $t = 2\pi/\omega_{eigen}$, measured in the oscillator periods. For example, $t = 1000$ corresponds to 1000 'attempts' to escape from the left well. However, one can see that the probability to find the particle in the right well is still negligibly small. Only at *t =* 10000, it becomes higher than the probability to find the particle in the left well.

The situation become dramatically different in the case of time-periodically driven wells [https://www.dropbox.com/s/eczwm6ny8f0939t/Potential%20driving.gif?dl=0], as demonstrated in Fig. 5 for the two driving frequencies $\Omega = \omega_{eigen}; 2\omega_{eigen}$. In both cases, already at t = 100, the probability to find the particle in the right well becomes comparable with the probability to find the particle in the left well. This means that the mean escape (tunneling) time has decreased by ~ **two orders of magnitude** due to the driving with a comparatively small driving amplitude $g_\omega = 2\, g_x = 0.1 \ll 1$.

Fig. 6 demonstrates dependence of the driving effect on the driving frequency, which is different from that obtained for a harmonic oscillator [13], where two sharp peaks were observed at resonant frequencies $\Omega = \omega_{eigen}$ and $\Omega = 2\omega_{eigen}$. Due to a *simultaneous* time-periodic modulation of the potential well *positions* and *eigenfrequencies*, the accelerating effect of driving depends non-monotonously on the driving frequency with a several maximums lying between $\omega_{eigen}$ and $2\omega_{eigen}$.

Finally, dependence of the tunneling time on the driving amplitude is shown in Fig. 7. It appears that increasing the amplitude by a factor of 2 results in decreasing the mean tunneling time by an order of magnitude. This example demonstrates the importance of the time-periodic driving of the potential wells in the vicinity of LAVs and phasons in the reactions involving quantum tunneling.

In the following section, we consider some characteristic examples of LAVs and phasons quasicrystals.

**Non-resonant driving with various frequencies $\Omega$**

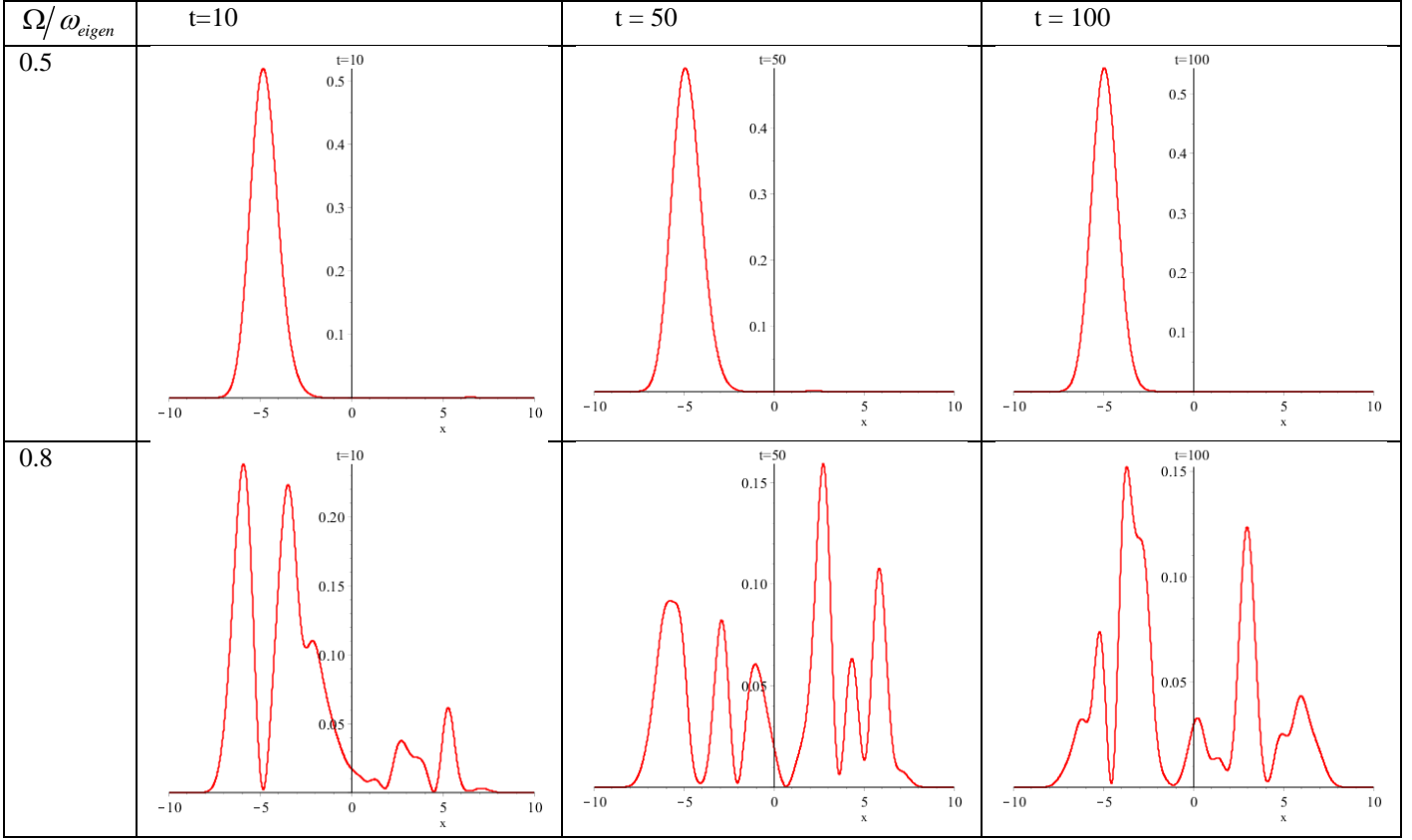

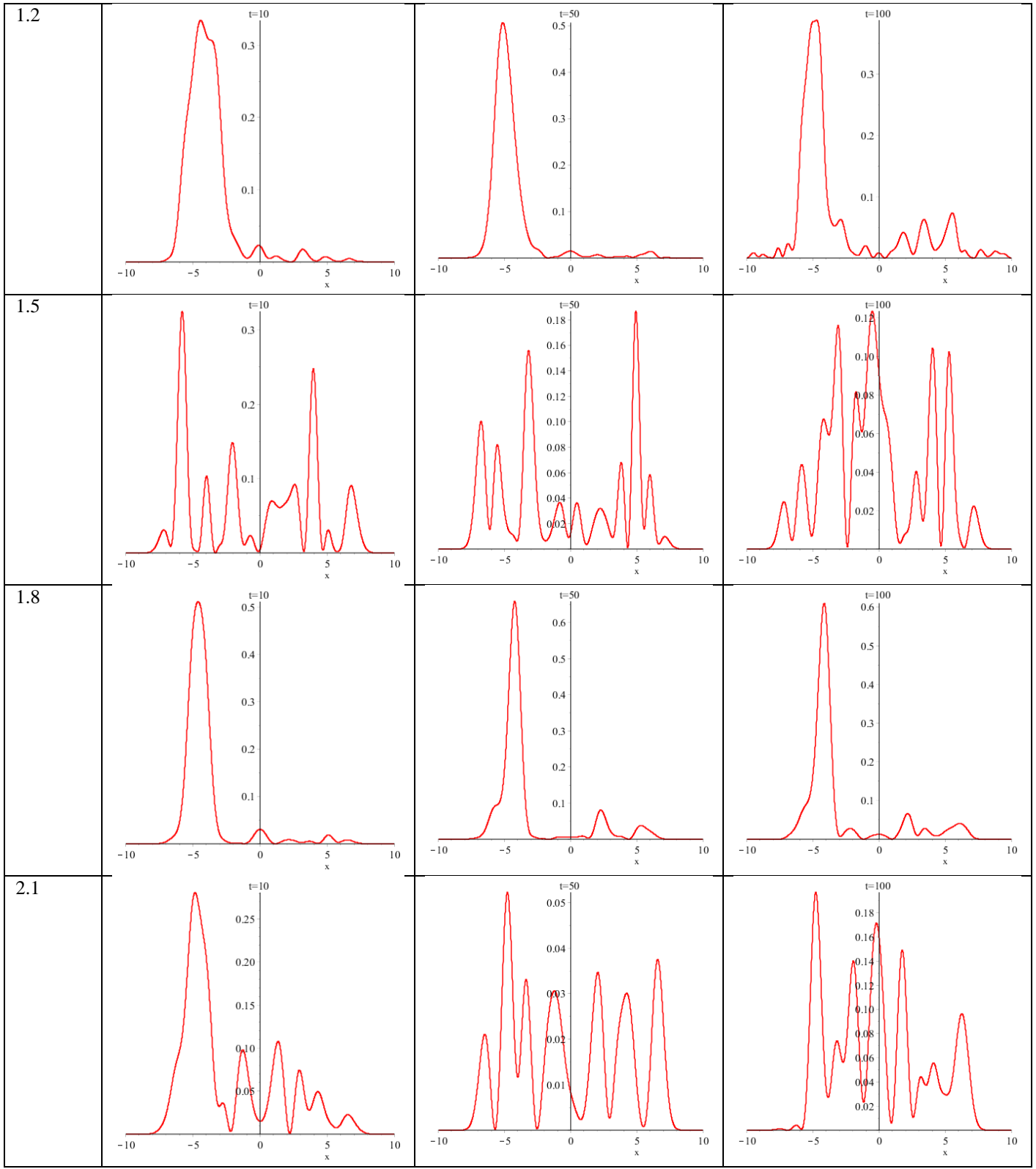

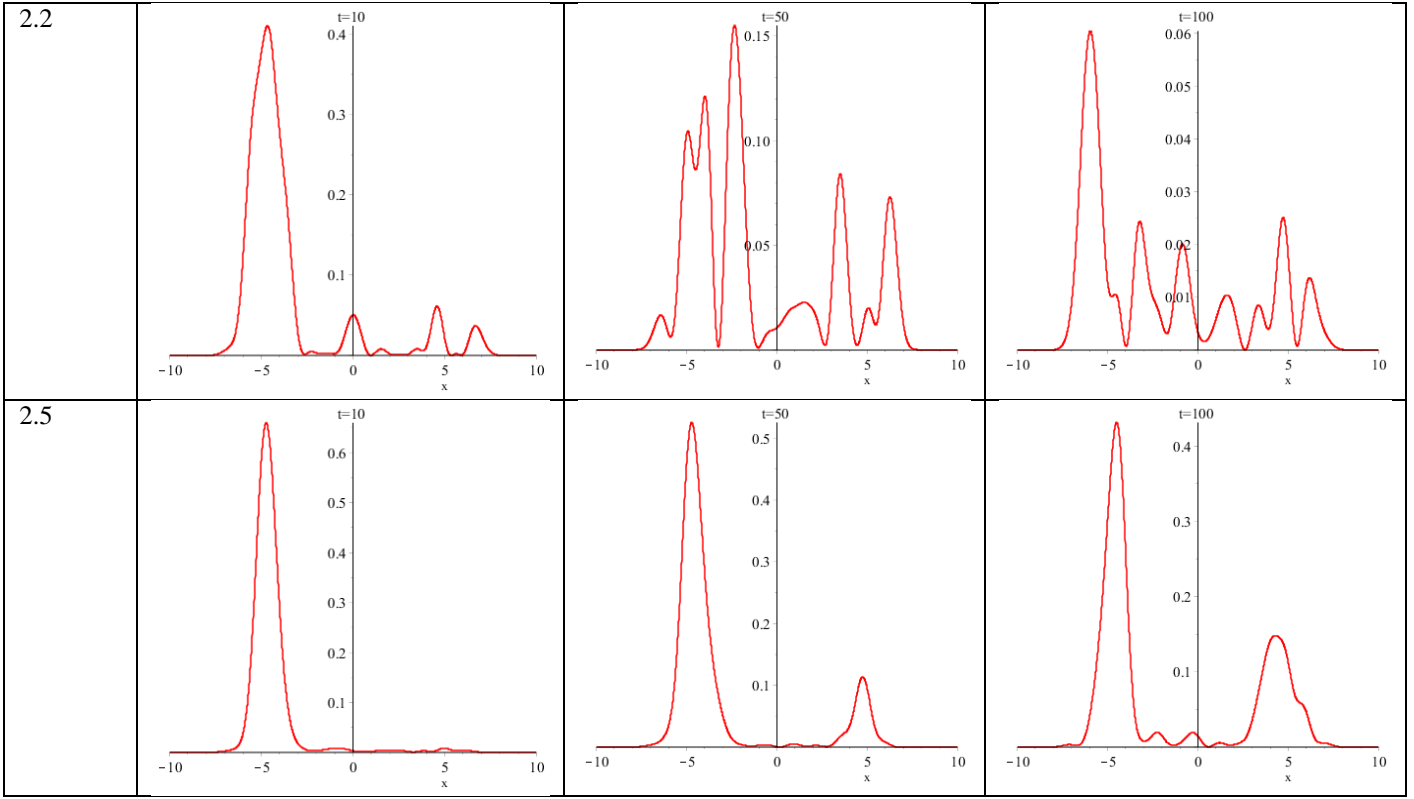

Figure 6. The probability distribution of the particle at different moments of time under the potential driving ($\alpha = 0.0005$; $\beta = 0.0001$) corresponding to $g_\omega = \beta/2\alpha = 0.1$; $g_x = \beta/4\alpha = 0.05$. The driving frequency $\Omega/\omega_{eigen}$ = 0.5; 0.8; 1.2; 1.5; 1.8; 2.1; 2.2; 2.5 is indicated in the figure.

**Effect of the driving strength/amplitude, g**

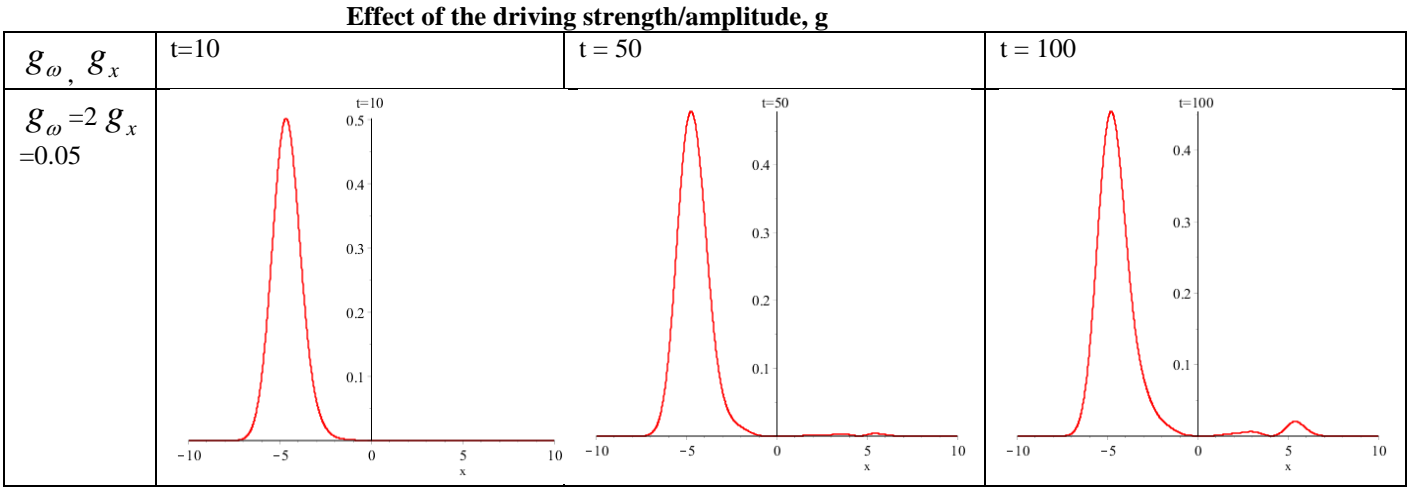

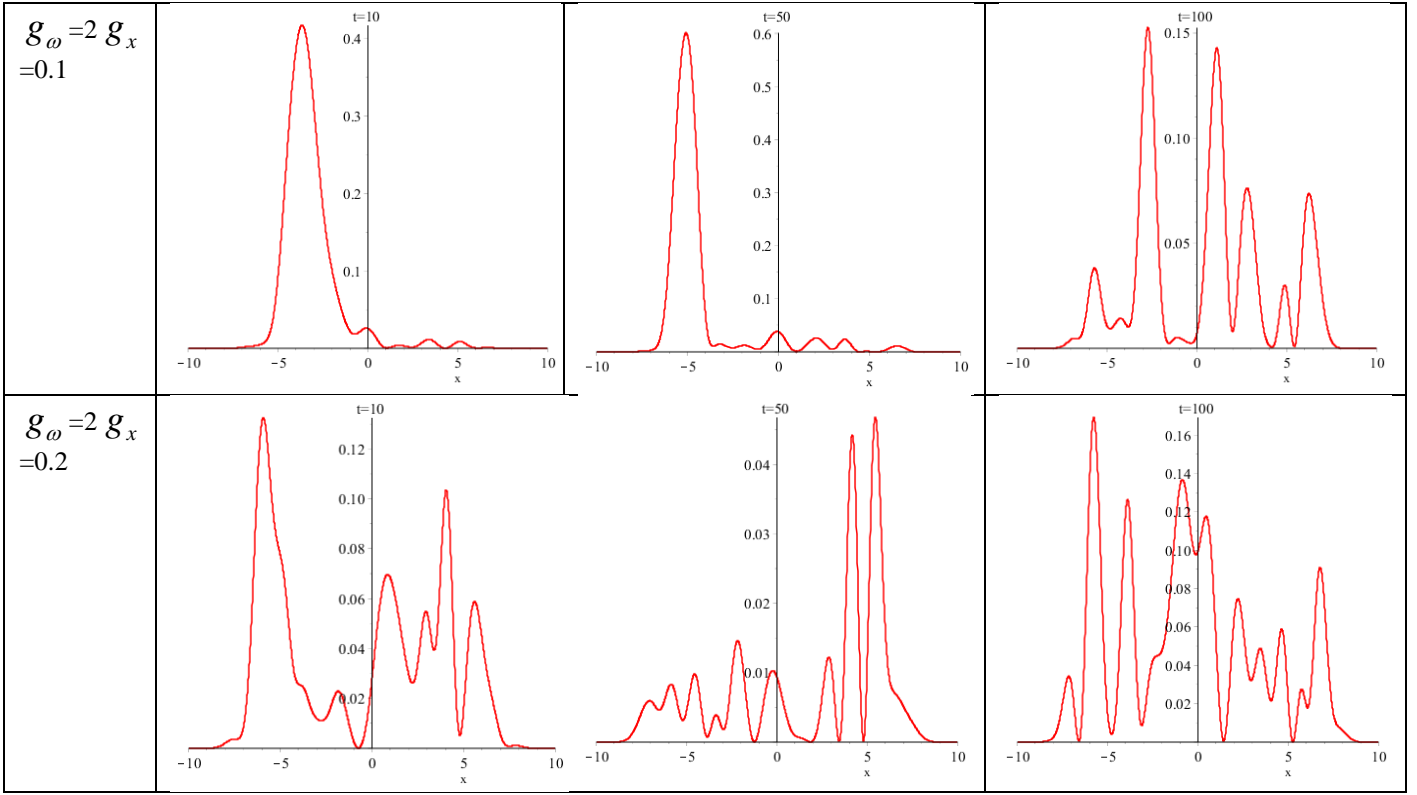

Figure 7. The probability distribution of the particle at different moments of time under the potential driving at $\Omega = \omega_{eigen}$, $\alpha = 0.0005$; $\beta = 0.00005 \div 0.0002$, corresponding to different driving amplitudes $g_\omega$, $g_x$ as indicated in the figure.

## 4. LAVs and phasons in nanocrystals and quasicrystals

### 4.1 LAVs in nanocrystals and quasicrystals

The fact that the energy localization manifested by LAV does not require long-range order was first realized as early as in 1969 by Alexander Ovchinnikov who discovered that localized long-lived molecular vibrational states may exist already in simple molecular crystals ($H_2$, $0_2$, $N_2$, NO, CO) [14]. He realized also that stabilization of such excitations was connected with the *anharmonicity* of the intramolecular vibrations. Two coupled anharmonic oscillators described by a simple set of dynamic equations demonstrate this idea:

$$\ddot{x}_1 + \omega_0^2 x_1 + \varepsilon \lambda x_1^3 = \varepsilon \beta x_2 \\ \ddot{x}_2 + \omega_0^2 x_2 + \varepsilon \lambda x_2^3 = \varepsilon \beta x_1 \quad , \tag{12}$$

where $x_1$ and $x_2$ are the coordinates of the first and second oscillator, $\omega_0$ are their zero-point vibrational frequencies, $\varepsilon$ is a small parameter, and $\lambda$ and $\beta$ are parameters characterizing the *anharmonicity* and the *coupling* force of the two oscillators, respectively. If one oscillator is displaced from the equilibrium and start oscillating with an initial amplitude, *A*, then the time needed for its energy to transfer to another oscillator is given by the integral:

$$T = \frac{\omega_0}{\varepsilon \beta} \int_0^{\pi/2} \frac{d\varphi}{\sqrt{1-\left(A^2\gamma/4\right)^2 \sin^2\varphi}}, \qquad \gamma = \frac{3\lambda}{\beta}, \tag{13}$$

from which it follows that the full exchange of energy between the two oscillators is possible only at sufficiently small initial amplitude: $A^2\gamma/4 < 1$. In the opposite case, $A^2\gamma/4 > 1$, the energy of the first oscillator will *always be larger* than that of the second one. And for sufficiently large initial amplitude, $A \gg \sqrt{4/\gamma}$, there will be practically no sharing of energy, which will be localized exclusively on the first oscillator.

Thus, Ovchinnikov has proposed the idea of LAV for molecular crystals, which was developed further for any nonlinear systems possessing *translational symmetry*; in the latter case, LAVs have been named *discrete breathers* (DBs) or *intrinsic localized modes* (ILMs). Now, we are coming back to the idea of LAV arising at '*active sites*' in defected crystals, quasicrystals and nanoclusters. As noted by Storms, 'Cracks and small particles are the Yin and Yang of the cold fusion environment'. A physical reason behind this phenomenology is that in topologically disordered systems, sites are not equivalent and band-edge phonon modes are intrinsically localized in space. Hence, different families of LAV may exist, localized at different sites and approaching different edge normal modes for vanishing amplitudes [9]. Thus, in contrast to perfect crystals, which produce DBs homogeneously, there is a striking *site selectiveness* of energy localization in the presence of spatial disorder, which has been demonstrated by means of atomistic simulations in biopolymers [9], metal nanoparticles [15] and, *experimentally*, in a decagonal *quasicrystal* $Al_{72}Ni_{20}Co_8$ [11].

The crystal shape of the nanoparticles (cuboctahedral or icosahedral) is known to affect their catalytic strength [16], and the possibility to control the shape of the nanoparticles using the amount of hydrogen gas has been demonstrated both experimentally by Pundt et al [17], and by means of atomistic simulations by Calvo et al [18]. They demonstrated that above room temperature the *icosahedral phase* should remain stable due to its higher entropy with respect to cuboctahedron (Fig. 8). And icosahedral structure is one of the forms quasicrystals take, therefore one is tempted to explore further the *link between nanoclusters and quasicrystals*.

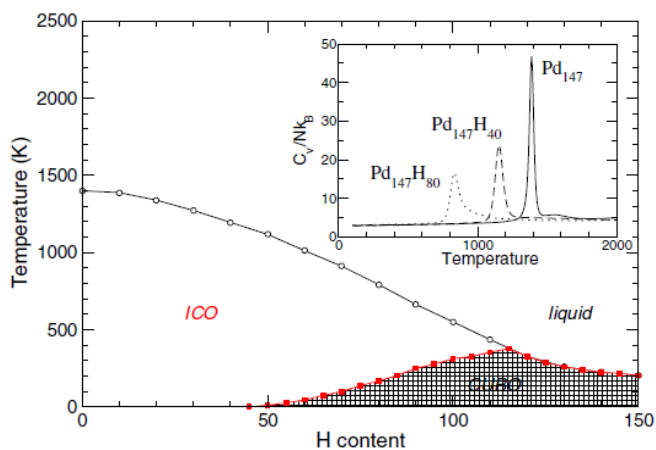

Figure 8. Schematic structural diagram of the $Pd_{147}H_x$ cluster in the icosahedral, cuboctahedral and liquid phases, after [18]. Inset: heat capacities of three clusters, in units of $k_B$ per atom, versus canonical temperature. Icosahedral phase is predicted to be more stable above room temperature.

Fig. 9 shows the structure of $Pd_{147}H_{138}$ cluster containing 147 Pd and 138 H atoms having minimum free energy configuration, replicated using the method and parameters by Calvo et al [18]. In particular, Fig. 9(**b**) reveals the presence of H-H-H chains aligned along the I-axis of the cluster. This *ab initio* simulation points out at the possibility of excitation of LAVs in these chains, with a central atom performing large-amplitude anharmonic oscillations and its neighbors oscillating in quasi-harmonic regime [19], which is similar to that considered in [7] for regular diatomic lattice of NaCl type. Such oscillations have been argued to facilitate LENR [2, 3], and in the present paper we develop this concept further.

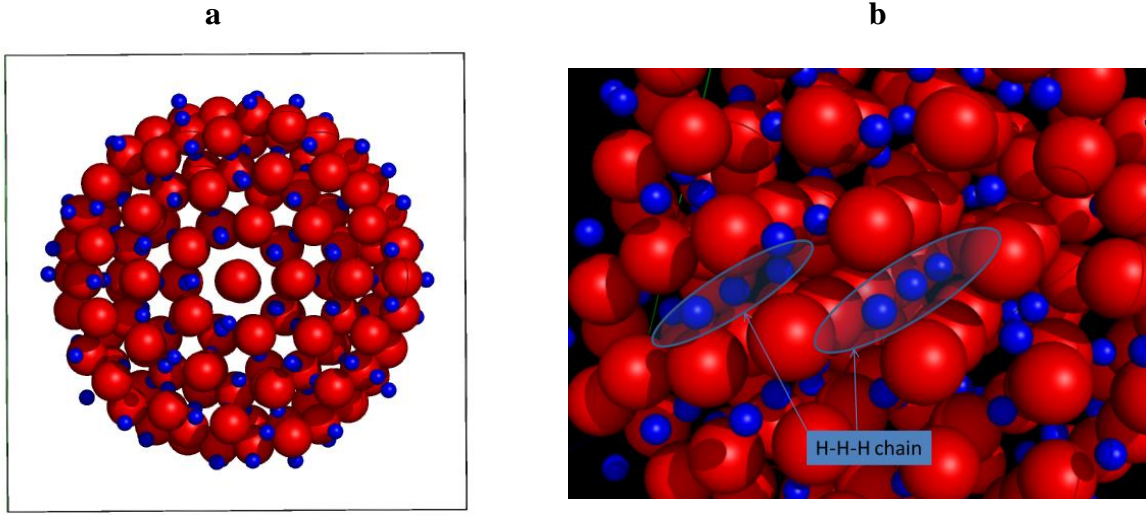

Figure 9. (**a**) Structure of PdH cluster containing 147 Pd and 138 H atoms having minimum free energy configuration, replicated using the method and parameters by Calvo et al [18]; (**b**) H-H-H chains in the nanocluster, which are viable sites for LAV excitation [19].

In the following section, we will consider phasons observed in a decagonal *quasicrystal* $Al_{72}Ni_{20}Co_8$ [11] and a possible link between LAVs and phasons.

*4.2 LAVs vs. phasons in quasicrystals*

Abe et [11] has measured by means of high resolution scanning transmission microscope (STEM) temperature dependence of the so-called Debye–Waller (DW) factor in decagonal *quasicrystal* $Al_{72}Ni_{20}Co_8$. DW factor is determined by the mean-square vibration amplitude of the atoms. The vibrations can be of thermal or quantum nature depending on the temperature. The authors demonstrated that the anharmonic contribution to Debye–Waller factor increased with temperature much stronger than the harmonic (phonon) one. This was the first underline{direct observation} of a 'local thermal vibration anomaly' i.e. LAVs, in our terms (Fig. 10). The experimentally measured separation between LAVs was about 2 nm, which meant that their mean concentration was about $10^{20}$ per $cm^3$ that is many orders of magnitude higher than one could expect to find in periodic crystals [7].

The LAV amplitude dependence on temperature fitted by two points at 300 K and 1100 K has shown that the maximum LAV amplitude at 1100K = 0.018 nm (Fig. 11**a**). What is more, it appears that LAVs give rise to phasons at T > 990 K, where a phase transition occurs, and additional quasi-stable sites β arise near the sites α. The phason amplitude of 0.095 nm (Fig. 11**b**) is an ***order of magnitude larger*** than that of LAVs. Thus, on the one hand, the driving amplitude induced by phasons is larger than that by LAVs, but on the other hand, phason oscillations may be less time-periodic (more stochastic), which requires more detailed investigations of the driving stochasticity effect on tunneling, as discussed in the following section.

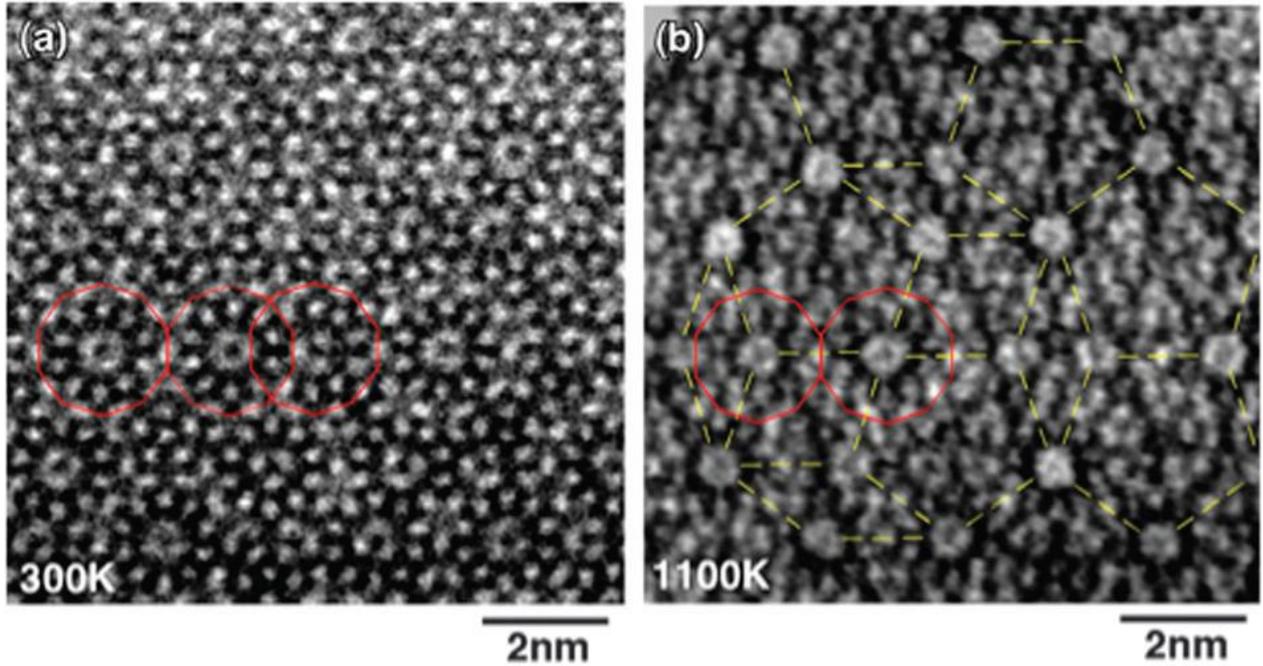

Figure 10. STEM images of LAVs of the decagonal $Al_{72}Ni_{20}Co_8$ at (a) 300 K and (b) 1100 K, according to Abe et al[11]. Connecting the center of the 2 nm decagonal clusters (red) reveals significant temperature-dependent contrast changes, a pentagonal quasiperiodic lattice (yellow) with an edge length of 2 nm can be seen in (b).

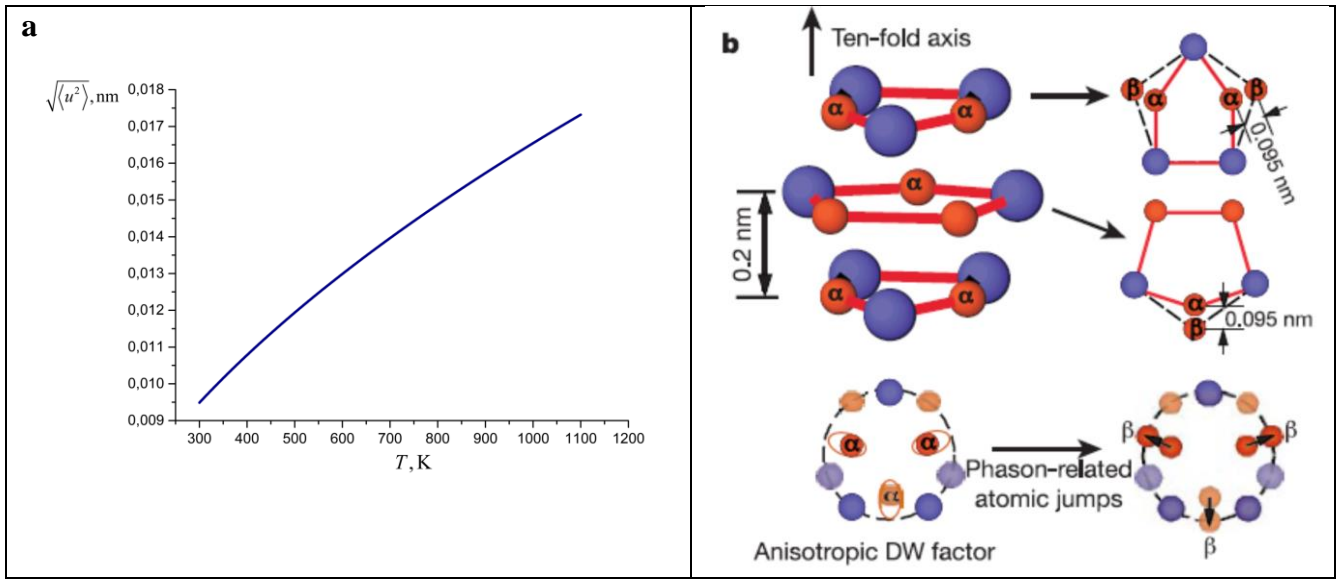

Figure 11. (a) LAV amplitude dependence on temperature in $Al_{72}Ni_{20}Co_8$, fitted by two points at 300 K and 1100 K, according to Abe et al [11]. The maximum LAV amplitude at 1100K = 0.018 nm.
(b) LAVs give rise to phasons at T > 990 K, where a phase transition occurs, and additional quasi-stable sites β arise near the sites α. The phason amplitude of 0.095 nm is an order of magnitude larger than that of LAVs.

## 5. Conclusions and outlook

In the present paper, we presented numerical solution of Schrodinger equation for a particle in a *nonstationary* double well potential, which is driven *time-periodically* imitating the action of a LAV or a phason on the reaction cite in their vicinity. We have shown that the rate of tunneling of the particle through the potential barrier separating the wells can be enhanced by *orders of magnitude* with increasing number of driving periods. This effect is novel, since it differs qualitatively from a well-studied effect of resonance tunneling [20-24], a.k.a. Euclidean resonance (an easy penetration through a classical nonstationary barrier due to an under-barrier interference) [20-23]. In the latter case, the tunneling rate has a sharp peak as a function of the particle energy when it is close to the certain *resonant value* defined by the nonstationary field. Therefore, it requires a very specific parametrization of the tunneling conditions. In contrast to that, the time-periodic driving of the potential wells considered above, results, first of all, in a sharp and continuous (not quantum) increase of the ZPV energy [6, 13], which in its turn increases the tunneling rate. It increases strongly with increasing strength of the driving, which is related to the amplitude of the non-linear dynamic phenomenon that causes the driving. As we have demonstrated in the previous section, the driving amplitude induced by phasons may larger than that by LAVs by an order of magnitude, which implies that phasons may be stronger catalysts than LAVs. However, further research is needed in order to make more definite conclusions, since the phason dynamics itself is an activated process driven by thermal or quantum fluctuations. Therefore, phasons can hardly induce a strictly time-periodic driving considered in the present paper. Tunneling rate through a fluctuating barrier in the presence of a periodically driving field has been shown to decrease with increasing fluctuation strength [24]. One may expect similar effects due to fluctuations in the cases of LAV and phason driven tunneling, which requires further investigations.

In conclusion, the present results support the concept *of nuclear catalysis* in QCs that can take place at special sites provided by their inherent topology, which makes QCs a promising nuclear active environment.

**Acknowledgements**

The authors would like to thank Dmitry Terentyev for designing Fig. 8 and Dan Woolridge – LAV animation [19]. VD and DL gratefully acknowledge financial support from Quantum Gravity Research.